 \documentclass[iop]{emulateapj}

\usepackage{amsmath}

\newcommand{\myemail}{ricardo.amorin@oa-roma.inaf.it}

\newcommand{\hii}{\relax \ifmmode {\mbox H\,{\scshape ii}}\else H\,{\scshape ii}\fi}
\newcommand{\mi}{\relax \ifmmode {\mu{\mbox m}}\else $\mu$m\fi}
\newcommand{\ha}{\relax \ifmmode {\mbox H}\alpha\else H$\alpha$\fi}
\newcommand{\hb}{\relax \ifmmode {\mbox H}\beta\else H$\beta$\fi}
\newcommand{\hg}{\relax \ifmmode {\mbox H}\beta\else H$\gamma$\fi}
\newcommand{\hd}{\relax \ifmmode {\mbox H}\beta\else H$\delta$\fi}

\newcommand{\sii}{\relax \ifmmode {\mbox S\,{\scshape ii}}\else S\,{\scshape ii}\fi}
\newcommand{\siii}{\relax \ifmmode {\mbox S\,{\scshape iii}}\else S\,{\scshape iii}\fi}
\newcommand{\nii}{\relax \ifmmode {\mbox N\,{\scshape ii}}\else N\,{\scshape ii}\fi}
\newcommand{\oii}{\relax \ifmmode {\mbox O\,{\scshape ii}}\else O\,{\scshape ii}\fi}
\newcommand{\oiii}{\relax \ifmmode {\mbox O\,{\scshape iii}}\else O\,{\scshape iii}\fi}
\newcommand{\neiii}{\relax \ifmmode {\mbox Ne\,{\scshape iii}}\else Ne\,{\scshape iii}\fi}
 
 \newcommand{\rdostres}{\relax \ifmmode {\,\mbox{R}}_{\rm 23}\else \,\mbox{R}$_{\rm 23}$\fi}

\slugcomment{To appear in {\it The Astrophysical Journal Letters}}

\shorttitle{A low-metallicity dwarf galaxy at z$\sim$\,3.5}
\shortauthors{R. Amor\'in et al.}

\begin{document}


\title{Evidence of very low metallicity and high ionization state in a
  strongly lensed, star-forming dwarf galaxy at
  z$=$\,3.417\thanks{Based on Large Binocular Telescope (LBT) observations}}

\author{
R. Amor\'in \altaffilmark{1},
A. Grazian\altaffilmark{1}, 
M. Castellano\altaffilmark{1},
L. Pentericci\altaffilmark{1}, 
A. Fontana\altaffilmark{1}, 
V. Sommariva\altaffilmark{1},
A. van der Wel\altaffilmark{2},
M. Maseda\altaffilmark{2},
E. Merlin\altaffilmark{1}
  }
\affil{(1) INAF -- Osservatorio Astronomico di Roma, via Frascati 33, 00040  Monteporzio Catone, Roma, Italy}
\affil{(2) Max-Planck-Institut f\"ur Astronomie, K\"onigstuhl 17, D-69117, Heidelberg, Germany}



\altaffiltext{1}{ASTRODEEP fellow; Email: \myemail}


\begin{abstract}
We investigate the gas-phase metallicity and Lyman Continuum (LyC) escape
  fraction of a strongly gravitationally lensed, extreme emission-line galaxy at
  $z=3.417$, $J1000+0221S$, recently discovered by the CANDELS
  team. We derive ionization and metallicity sensitive emission-line ratios
  from H+K band LBT/LUCI medium resolution spectroscopy. 
$J1000+0221S$ shows high ionization conditions, as
  evidenced by its enhanced [\oiii]/[\oii] and [\oiii]/\hb\
  ratios. Consistently, strong-line methods based on the 
  available line ratios suggest that $J1000+0221S$ is an extremely 
metal-poor galaxy, with a  metallicity of {12$+\log$(O/H)$<$7.44}
  ($Z$\,$<$\,0.05\,$Z_{\odot}$, placing it among the most
  metal-poor star-forming galaxies at $z\ga$\,3 discovered so far. 
In combination with its low stellar mass 
(2$\times$10$^{8}$\,M$_{\odot}$) and
  high star formation rate (5\,M$_{\odot}$\,yr$^{-1}$), the
  metallicity of $J1000+0221S$ is consistent with the extrapolation to
  low masses of the mass-metallicity relation traced by 
  Lyman-break galaxies at $z\ga$\,3, but it is {0.55} dex lower than
  predicted by the fundamental metallicity relation at
  $z\la$\,2.5. These observations suggest the picture of a rapidly growing 
galaxy, possibly fed by the massive accretion of pristine gas.  
 Additionally, deep LBT/LBC in the \textit{UGR} bands are used to
 derive a limit to the LyC escape fraction,  thus allowing us to explore for the 
first time the regime of sub-$L^{*}$ galaxies at $z>$\,3. 
We find a 1$\sigma$ upper limit to the escape fraction of 23\%,
which adds a new observational constraint to recent theoretical 
models predicting that sub-$L^{*}$ galaxies at high-$z$ have 
high escape fractions and thus are the responsible for the reioization of the
Universe.
\end{abstract}



\section{Introduction}
\label{s1}

Recently, \citet[][hereafter vdW13]{vanderWel2013} presented the
serendipitous discovery of the first strong galaxy lens at $z_{\rm
  lens}>$\,1, $J100018.47+022138.74$. 
This quadrupole lens system was found in the COSMOS field covered by  
the CANDELS \citep{Grogin2011,Koekemoer2011} survey. 
Using \textit{Hubble Space Telescope} (HST) near-infrared (NIR)
imaging from CANDELS and NIR spectroscopy from the Large Binocular 
Telescope (LBT), the authors reported a record lens redshift 
$z_{L}=$\,1.53\,$\pm$\,0.09 and a 
strongly magnified (40x) source at redshift 
$z_S=$\,3.417\,$\pm$\,0.001 (hereafter $J1000+0221S$). 
While the lens is a quiescent and relatively massive galaxy, the
magnified source was found to be a low-mass (M$_{\star}
\sim$\,10$^{8}$\,M${_{\odot}}$), extreme emission-line galaxy (EELG) 
with unusually high rest-frame [\oiii]$\lambda$\,5007\AA\ equivalent 
width ($EW_0 \sim 1000$\AA). 

The scarcity of strong galaxy lenses at high redshift makes the
discovery of $J100018.47+022138.74$ especially remarkable. 
Strikingly enough, the probability to find a EELG being lensed by
another galaxy appears to be very low, unless these galaxies become 
significantly more abundant at high-$z$ (vdW13). 
Consistently, a large number of low-mass EELGs at $z\sim$\,2 have 
started to emerge from deep surveys 
\citep[e.g.][]{vanderWel2011,Atek2011,Guaita2013,Maseda2013,Maseda2014} 
and recent observational evidences point to their ubiquitousness 
at $z\sim$\,5-7 \citep{Smit2013}. 
Low-mass galaxies with extreme nebular content 
at lower redshift are mostly chemically unevolved systems,
characterized by their compacteness, high \textit{specific} star 
formation rates (SFR), high ionization and low metallicities, which 
make them lie offset from the main sequence of galaxies 
in fundamental scaling relations between mass, metallicity 
and SFR \citep[e.g.][]{Amorin2010,Atek2011,Nakajima2013,Ly2014,Amorin2014b}. 
At high redshift, however, the full characterization of these
properties in intrinsically faint galaxies requires a enormous 
observational effort \citep[e.g.][]{Erb2010,Maseda2013} and 
detailed studies are mostly restricted to those sources being 
subject of strong magnification by gravitational 
lensing \citep[e.g.][]{Fosbury2003,
Richard2011,Christensen2012,Brammer2012,
Belli2013,Wuyts2012}. 

The aim of this \textit{Letter} 
is to fully characterize the lensed EELG 
$J1000+0221S$ at $z=3.417$ presented by vdW13. 
This unique galaxy will serve to investigate two key 
issues. 
Using the deepest available LBT photometry and  
spectroscopy we will first derive robust estimates of the 
ionization and metallicity properties of $J1000+0221S$ 
through strong emission line ratios. 
This provide additional hints on the evolutionary stage 
of the galaxy and allows to add new constrains to the 
low-mass end of the mass-metallicity-SFR relation 
at $z\sim$\,3.4.
Finally, $J1000+0221S$ will offer the opportunity to 
derive, for the first time, a limit to the Lyman Continuum 
(LyC) escape fraction at $z>3$ in the sub-$L^*$ regime, as 
suggested by \citet{vanzella12}.  

\begin{table}[t!]
\caption{Main derived properties of $J100018.47+022138.74$}
\label{Tab1}
\centering
\begin{tabular}{lc | lc }
\hline\hline
\noalign{\smallskip}
$RA$ ($J2000$) & 150.07697 & $z$ &	3.417  \\[3pt]
$DEC$ ($J2000$) & $+$2.36076&  [\oii]$\lambda\lambda$\,3727,3729/\hb  &$<$\,0.30   \\[3pt]
$M_{\rm B}$  & -17.8$\pm$0.3   &   [\oiii]$\lambda$\,4959/\hb & 1.44$\pm$1.35 \\[3pt]
$E(B-V)_{\star}$ &  $0.0\substack{+0.2 \\ -0.0}$ & [\oiii]$\lambda$\,5007/\hb & 4.47$\pm$1.25 \\[3pt]
$\log$\,M$_{\star}$ [M$_{\odot}$] & $8.41\substack{+0.25 \\ -0.30}$  &
 [\neiii]$\lambda$\,3868/\hb &$<$\,0.20  \\[3pt]
SFR [M$_{\odot}$\,yr$^{-1}$] & 5$\pm$2  & $12+\log({\rm O/H})$  & $<$\,$7.44\substack{+0.20 \\ -0.17}$ \\[3pt]
\noalign{\smallskip}							     
\hline							      			
\hline
\end{tabular}									
\begin{list}{}{}
\item Notes: {$B$-band absolute magnitude, stellar reddening, star
  formation rate and stellar mass were derived from the SED fitting
  after correction for magnification \citep{vanderWel2013}. Line fluxes are given relative to F(\hb)=1.} 
\end{list}
\end{table}

\section{Spectroscopic data and measurements}
\label{s2}

\begin{figure}[t!]
   \includegraphics[angle=0,width=8.65cm]{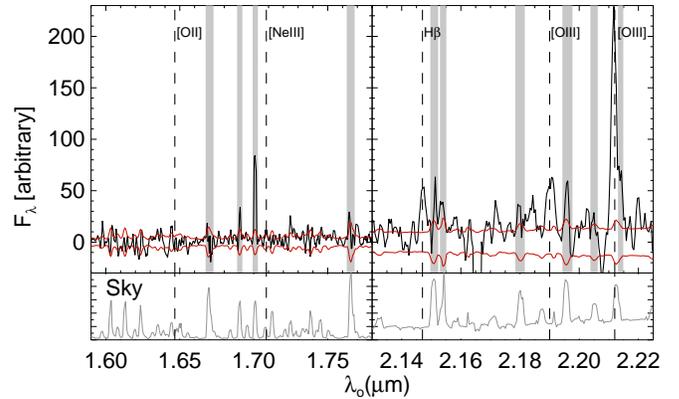}\hspace{5pt}
  \caption{LUCI/LBT $H+K$ spectrum of $J1000+0221$
    smoothed with the instrumental resolution. Overplotted in red we
    show the 3$\sigma$ noise spectrum. 
The position of both detected and undetected bright nebular
emission lines are indicated by dashed lines. 
Shaded vertical regions show spectral regions affected by strong sky emission lines.    
  }\label{fig:spectrum}
   \end{figure}

In order to derive line ratios for $J1000+0221S$, we re-analyse the 
LBT/LUCI NIR spectrum presented by vdW13 and 
shown in Fig.~\ref{fig:spectrum}. 
It consists of a 3-hour sequence of individual, dithered 120s
exposures using the H$+$K grism in a 1$"$ wide slit and seeing $\sim$0.6$"$. 
The spectrum was calibrated in wavelength and corrected by telluric
absorption, but an absolute flux calibration was not
possible (see vdW13). 
Full details on the reduction procedure are {discussed in a
  forthcoming paper \citet{Maseda2014}}.
 
Alternatively, we determine a relative flux calibration between the spectra coming from 
the two grisms, $H$ and $K$, after computing the integral flux of the two  
spectral ranges and their corresponding colour ($H-K$)$_{\rm spec}$. 
Then, we compare ($H-K$)$_{\rm spec}$ and the analogue colour, ($H-K$)$_{\rm phot}$,
derived from the {NEWFIRM} Medium Band Survey (NMBS) multiwavelength 
catalogue \citep{Whitaker2011}. 
We find the difference between the two measurements $\sim$\,10\%, so we conclude 
that uncertainties due to flux calibration in the derived flux ratios involving lines 
in the blue and red part of the spectrum are not significant.  

Integrated fluxes and uncertainties for the three emission 
lines detected in the $K$ spectrum, namely \hb, and 
[\oiii]\,$\lambda\lambda$4958,5007\AA, were measured 
using the IRAF task {\sl splot}, 
by summing all the pixels of the emission line after a linear 
subtraction of the continuum. 
Uncertainties are derived following \citet{Amorin2010}. 
We do not detect any line above 3$\sigma$ in the $H$ band spectrum. 
However, it is still possible to place stringent upper limits for [\oii] and 
[\neiii] emission lines as $F_{l}<$\,5$\sigma_{l}$, where $\sigma_l$ is the 
integral \textit{rms} noise of the spectrum over the total velocity width expected for 
the line. For the latter we assume the average width of the detected 
[\oiii] lines (13\AA).

Emission line ratios are computed after de-reddening the derived
fluxes and upper limits using the \citet{Calzetti2000} extinction law. 
Apart of \hb, no other Balmer lines are detected in the 
spectrum {due to sensitivity limits}. 
{Thus, based on the stellar reddening derived by vdW13, 
$E(B-V)_{\star}=0\substack{+0.2 \\  -0.0}$, we assume here 
$E(B-V)_{\rm gas}=0.2$ ($=1\sigma$ uncertainty in $E(B-V)_{\star}$), 
consistently with} the mean $E(B-V)_{\rm gas}$ found for low-mass 
galaxies of  similar EW[\oiii] in \citet[][]{Amorin2012,Amorin2014a,Amorin2014b}. 
Clearly, using a higher extinction we derive even more conservative
upper limits in metallicity. However, we note that a factor of 2
higher in extinction should {not significantly affect our results}. 

\section{Emission line and metallicity properties}
\label{s3}

\begin{figure*}[t!]
   \centering
   \includegraphics[angle=90,width=5.45cm]{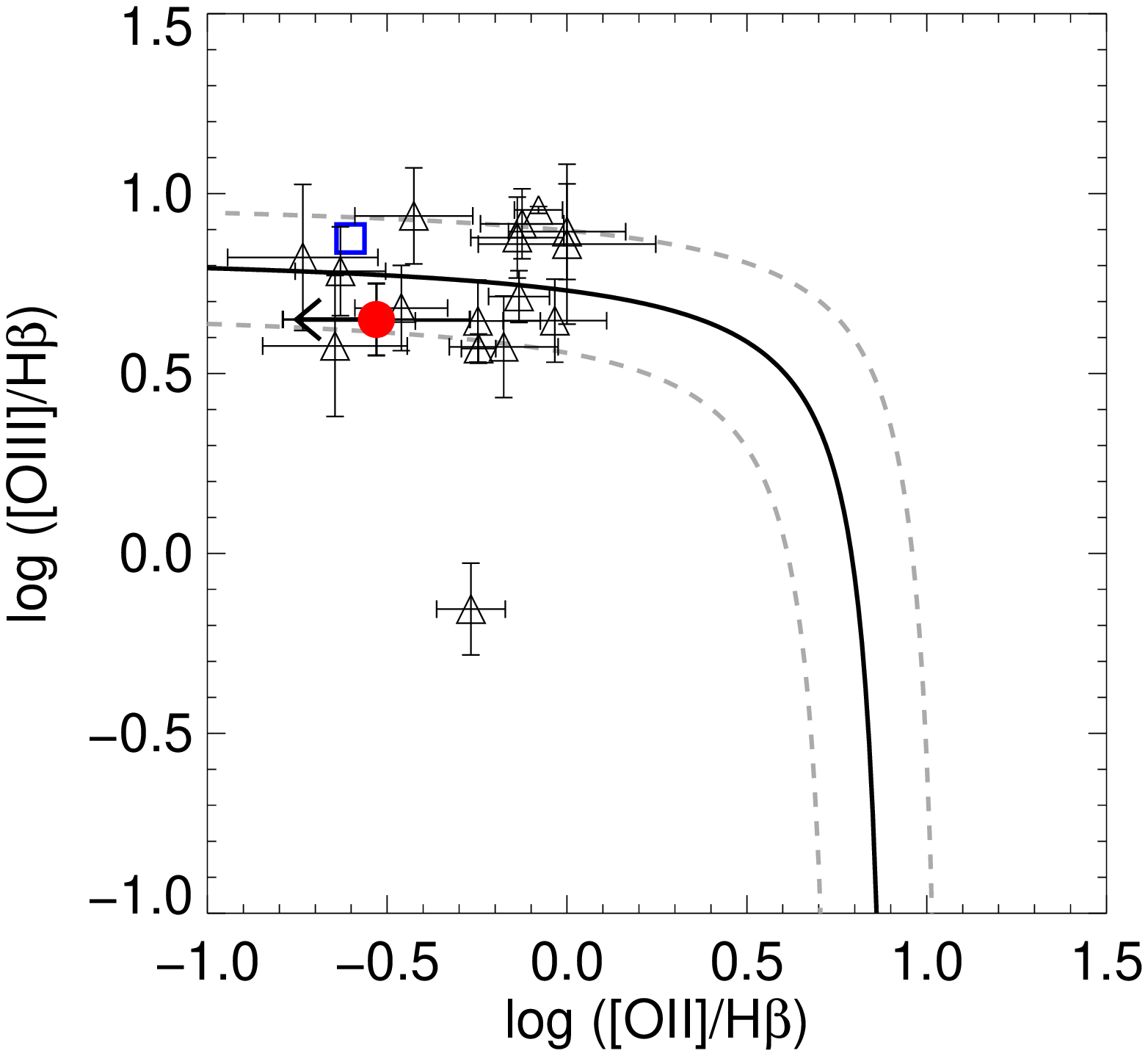} \hspace{1pt}
   \includegraphics[angle=90,width=5.3cm]{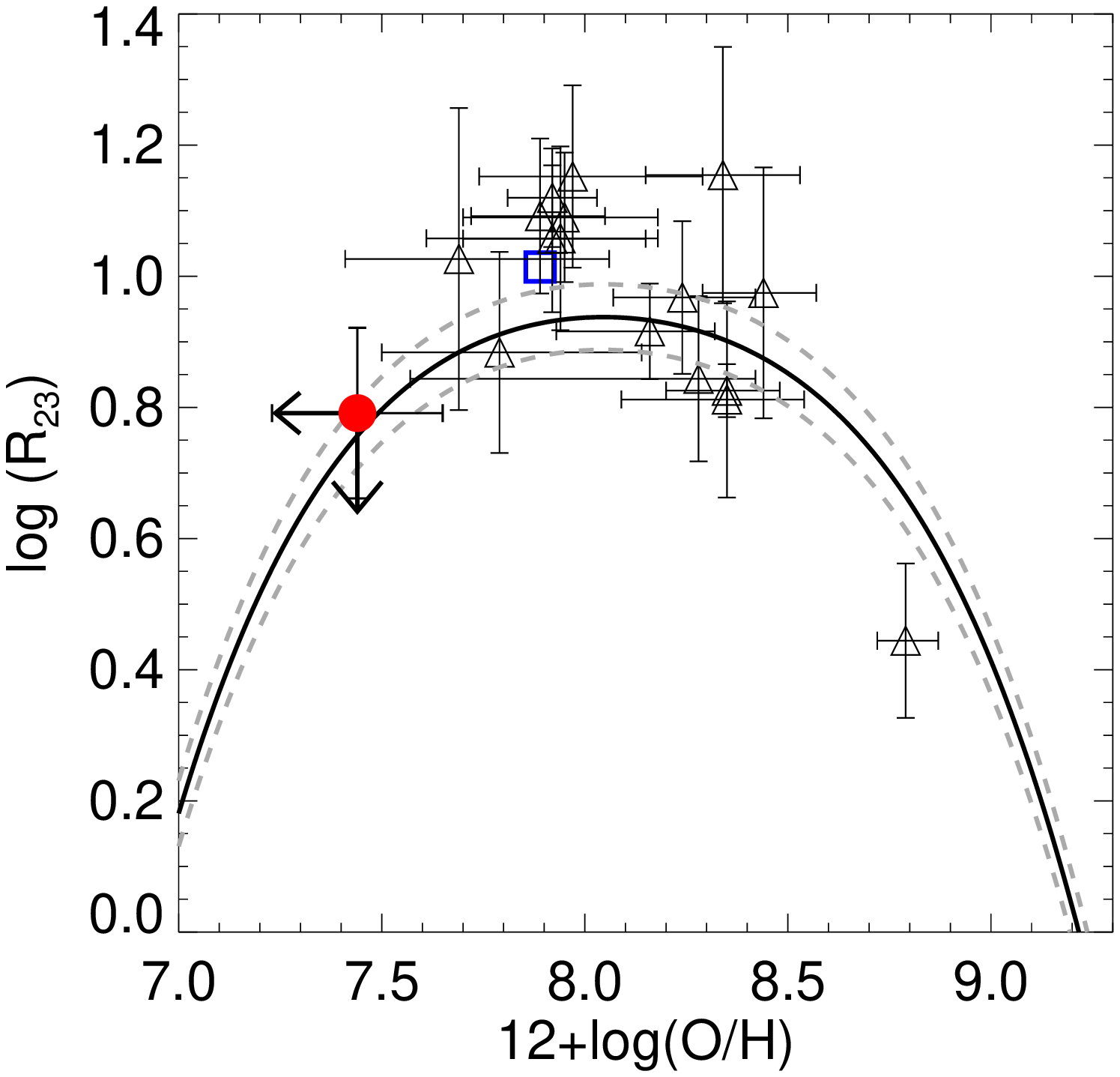} \hspace{1pt}
   \includegraphics[angle=90,width=5.5cm]{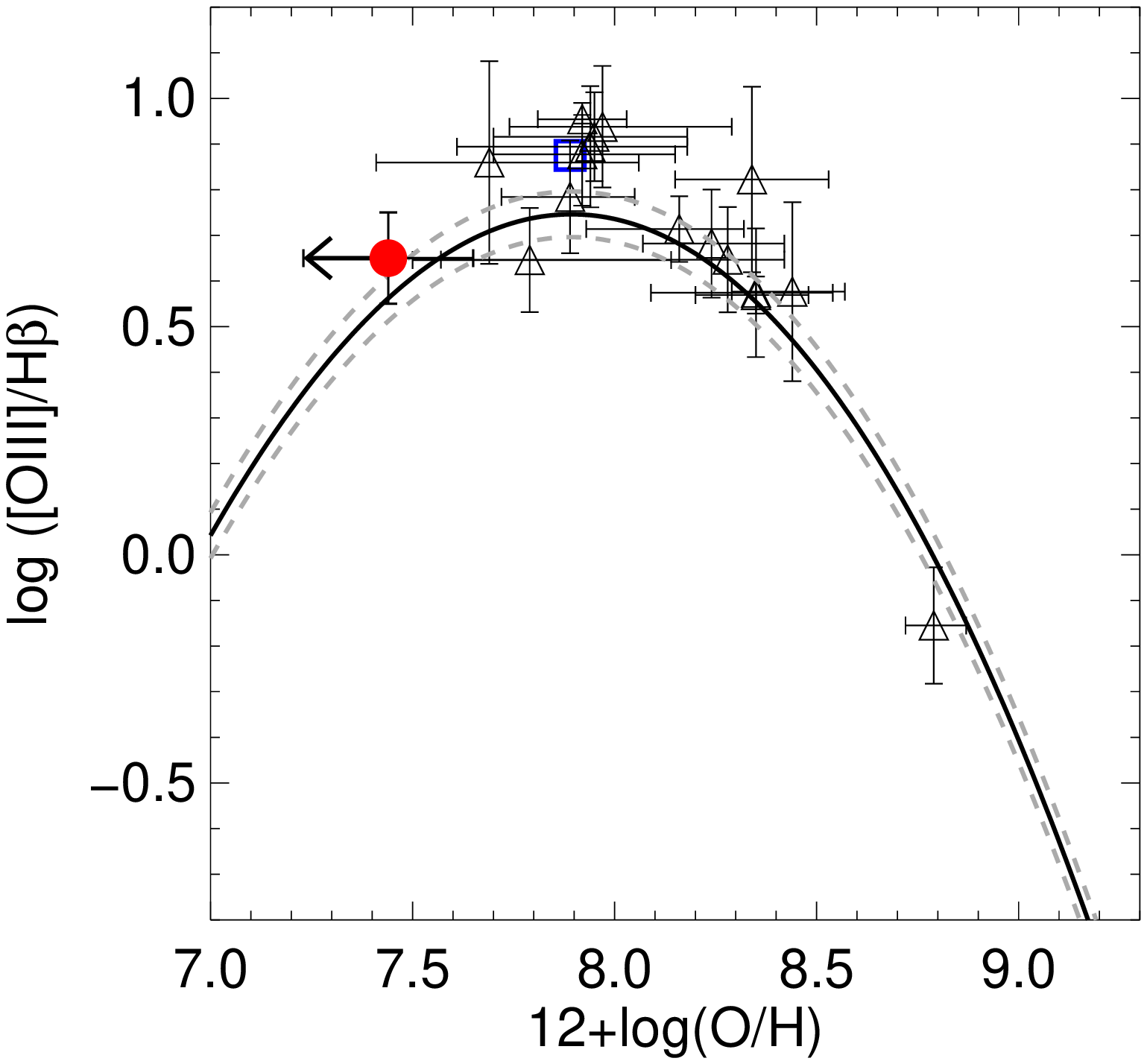} \\
   \includegraphics[angle=90,width=5.5cm]{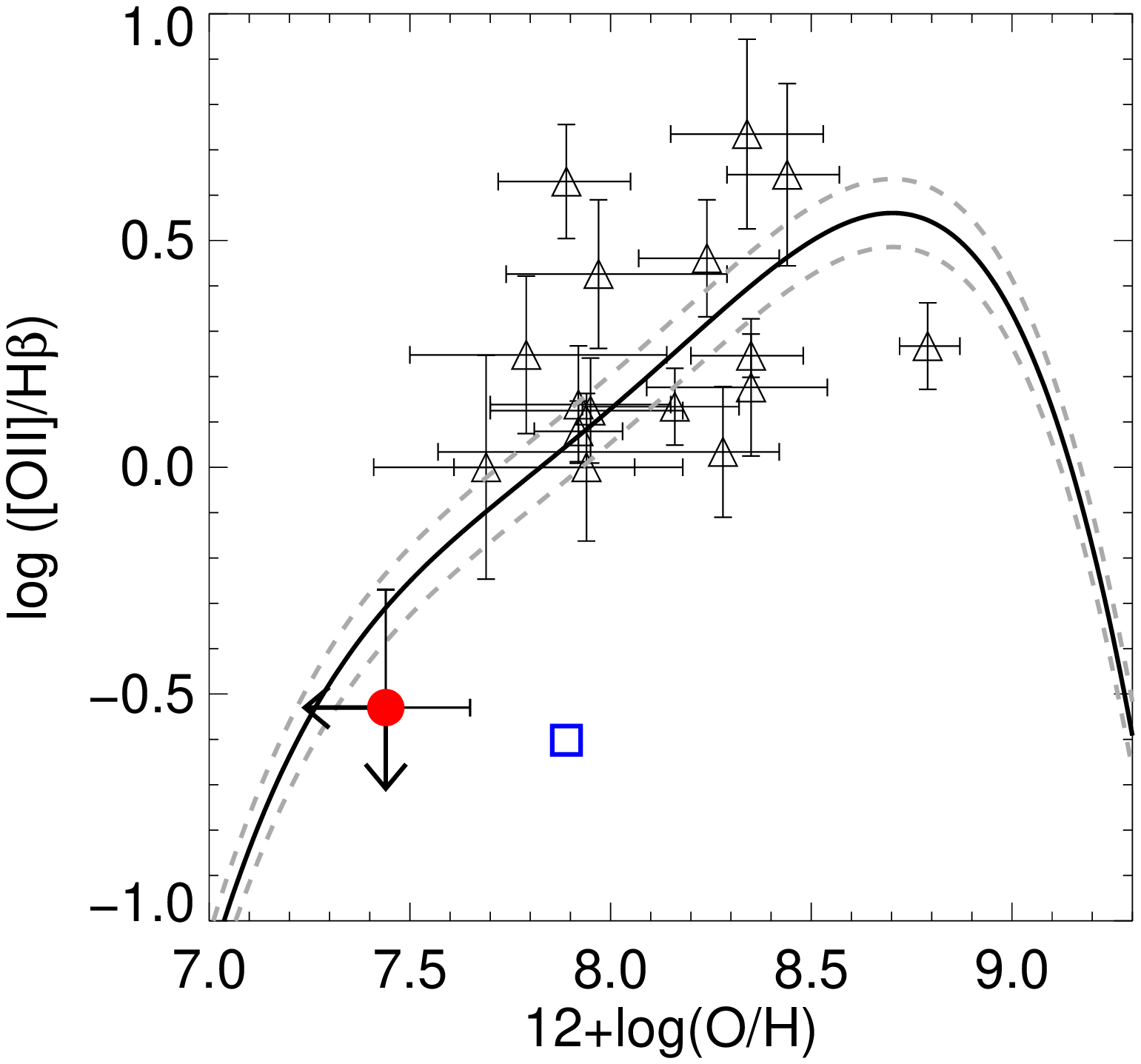} \hspace{1pt}
   \includegraphics[angle=90,width=5.5cm]{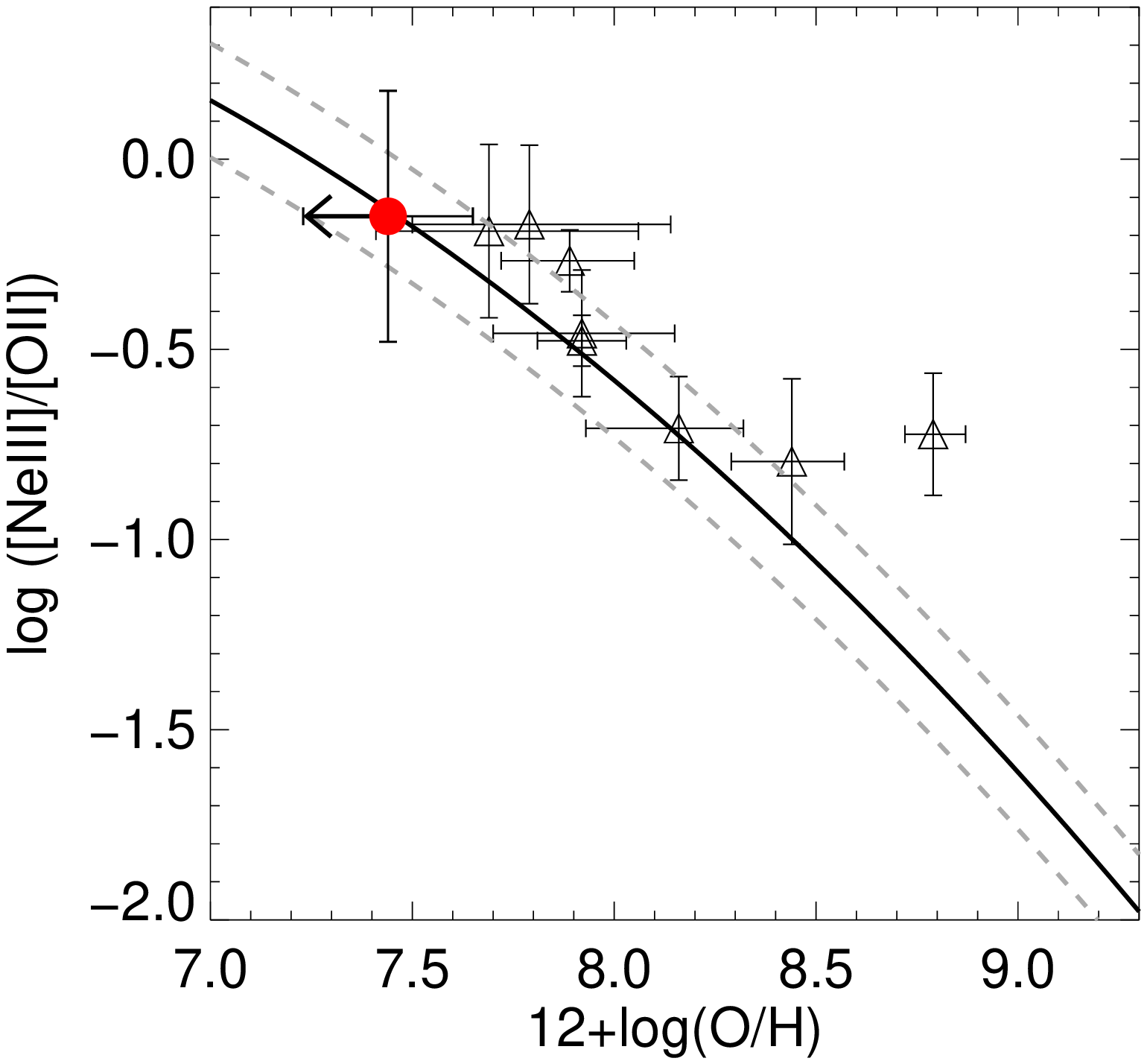} \hspace{1pt}
   \includegraphics[angle=90,width=5.5cm]{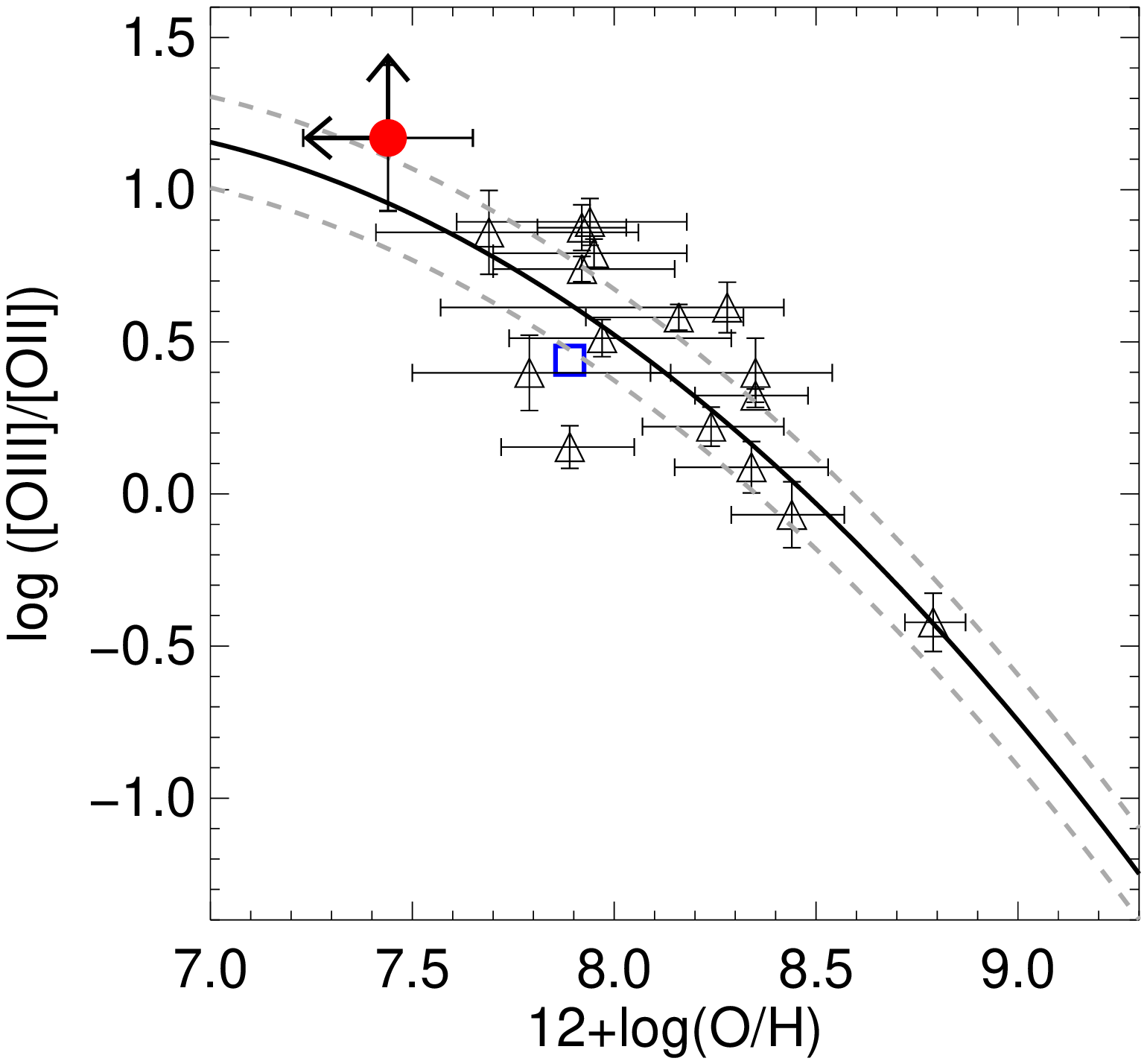} 
   \caption{Diagnostic diagrams and strong-line metallicity
     calibrations {from \citet{Maiolino2008} (see also Nagao et al., 2006). Grey dashed lines are 1$\sigma$ uncertainties for these relations. 
The best metallicity limit for $J1000+0221S$ that satisfy the various 
emission-line ratios is by adopting the lower $R23$ branch, as shown 
by the red circle.} LBGs at $z\geq$\,3 from the AMAZE
\citep{Maiolino2008} and LSD \citep{Mannucci2009} samples 
\citep[see also][triangles]{Troncoso2013} and the strongly lensed Lynx 
galaxy \citep[][blue square]{Fosbury2003} are shown for comparison.}\label{fig:calibraciones} 
     \end{figure*}

In Table~\ref{Tab1} we present the emission-line ratios and upper
limits relative to \hb\ derived for $J1000+0221S$, along with other 
physical properties taken from vdW13. 
The available line ratios are used in Fig.~\ref{fig:calibraciones} to
study the ionization and metallicity properties of $J1000+0221S$
through different diagnostic diagrams.  
In particular, the [\oiii]/\hb\ and [\oii]/\hb\ line
ratios suggest that $J1000+0221S$ is a high excitation, 
low-metallicity galaxy.  
Consistently, the remarkably low [\oii]/[\oiii] ratio suggests an
unusually high ionization. 
This is confirmed by the high ionization parameter, 
$\log q_{ion} > 8.5$\,cm\,s$^{-1}$, derived 
from the ionization-sensitive ratio [\oii]/[\oiii] and adopting 
the calibration based on photoionization models by 
\citet{Levesque2014}. 
Being higher than those found in normal galaxies, the derived 
$q_{ion}$ is only comparable to those found in other EELGs 
\citep[e.g.][]{Nakajima2013,Ly2014,Amorin2014a,Amorin2014b}. 

Strong telluric absorption 
{makes it impossible to detect the} faint $T_{e}-$sensitive 
auroral line [\oiii]$\lambda$4363\AA, precluding a direct
determination of the oxygen abundance. 
Alternatively, {we use the set of collisionally excited emission lines 
available for $J1000+0221S$ to derive its metallicity following 
a strong-line method based on the} 
$R23$ ($\equiv$([\oii]$+$[\oiii)]/\hb) {parameter, which
shows a well-known bi-valued relation with $Z$.  
In particular, here we use the calibration and the method proposed and
explained in detail by \citet{Maiolino2008}. The use of this
calibration give us the possibility of comparing our results with other 
LBGs at similar redshift \citep[e.g.][]{Troncoso2013}. }

{In short, this method combines the calibration of $R23$ and other 
nebular line ratios as a function of metallicity, as shown in 
Fig.~\ref{fig:calibraciones}. 
In the low metallicity regime ($Z\la$\,0.5$Z_{\odot}$) these
relations are calibrated using local galaxies with direct ($T_e$)
metallicities, while at higher $Z$ they are based on photoionization 
models. 
This method has the advantage that at least two of the relations  
involved increase monotonically with $Z$, so they can remove the
degeneracy in the $R23-Z$ relation. }
In particular, $J1000+0221S$ shows a remarkably low [\oii] upper 
limit, which makes any diagnostic based on such line, 
e.g. [\oii]/\hb\ vs. $Z$ or [\oiii]/[\oii] vs. $Z$ 
in Fig.~\ref{fig:calibraciones}, almost conclusive about this object
being extremely metal-poor. 
{In practice, we solve the calibrations shown in 
Fig.~\ref{fig:calibraciones} simultaneously and iteratively. 
An uncertainty of a factor of 2 in the adopted E(B-V)$_{\rm gas}$ 
has been propagated to the error bars shown in Fig.~\ref{fig:calibraciones}. 
Finally, the best metallicity limit that satisfy the various
emission-line ratios is by adopting the lower $R23$ branch, resulting
in an upper limit in metallicity 12$+\log$(O/H)$<$7.44$\pm$0.2 
at 68\% confidence level. }

\section{Discussion}
\label{s4}
\subsection{The low-mass end of the mass metallicity relation at
  z$\sim$3.4} 
\label{s4.1}

At redshift $z \sim$\,2-3, where massive galaxies are rapidly 
assembling most of their present-day stellar mass 
\citep[][]{Hopkins2006}, the 
 mass-metallicity relation (MZR) traced 
by luminous LBGs shows significant evolution 
\citep{Erb2006,Maiolino2008,Mannucci2009,Troncoso2013}. 
However, due to the challenge that measuring metallicities 
represents and poor statistics, the shape and normalization of the 
MZR at $z\ga$\,3 are still poorly constrained, especially in its 
low-mass end. Thus it is particularly interesting to study the 
position of $J1000+0221S$ in the MZR. 

\begin{figure}[t!]
   \centering
   \includegraphics[angle=90,width=8.7cm]{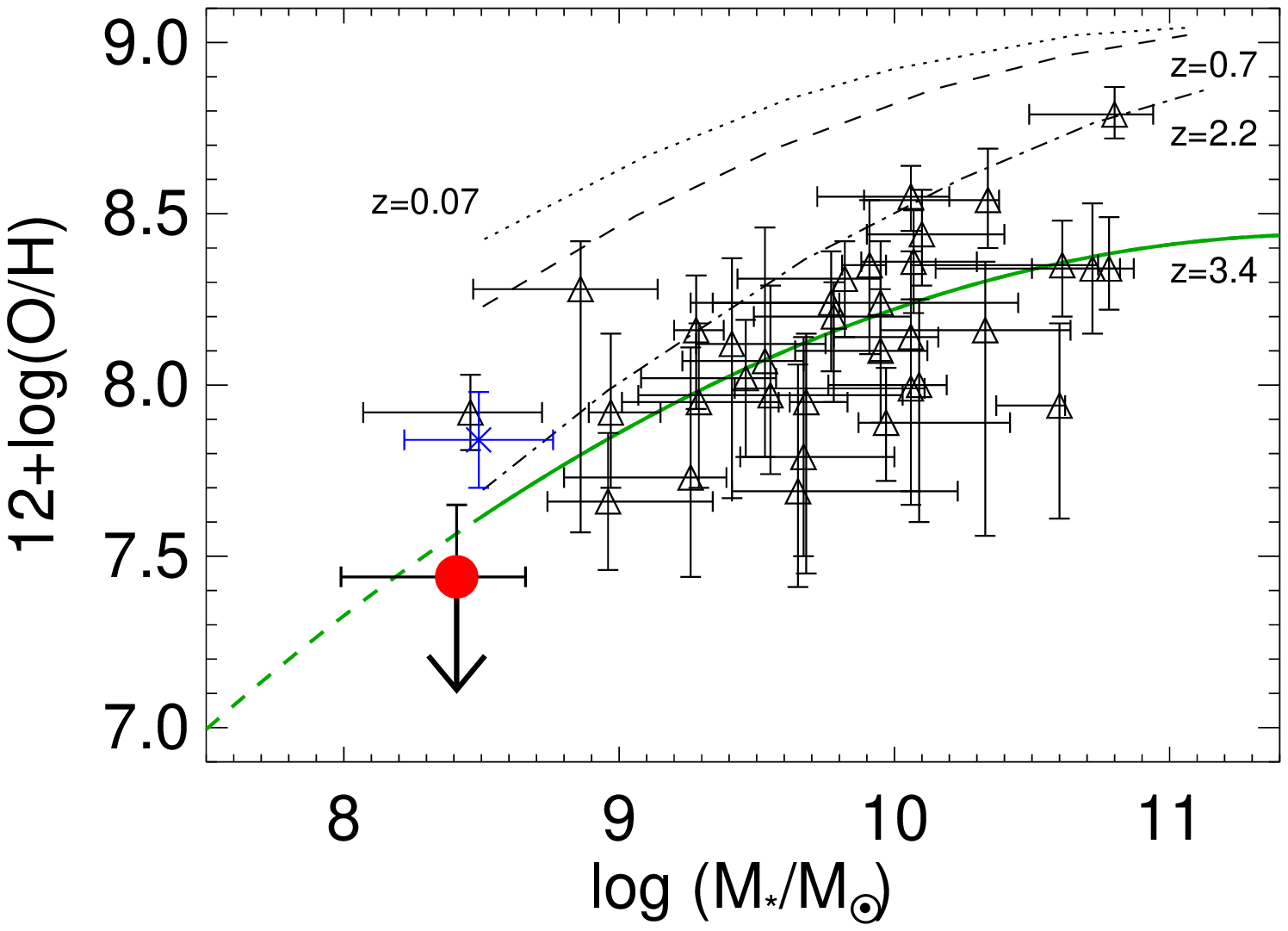} \\
   \includegraphics[angle=90,width=8.7cm]{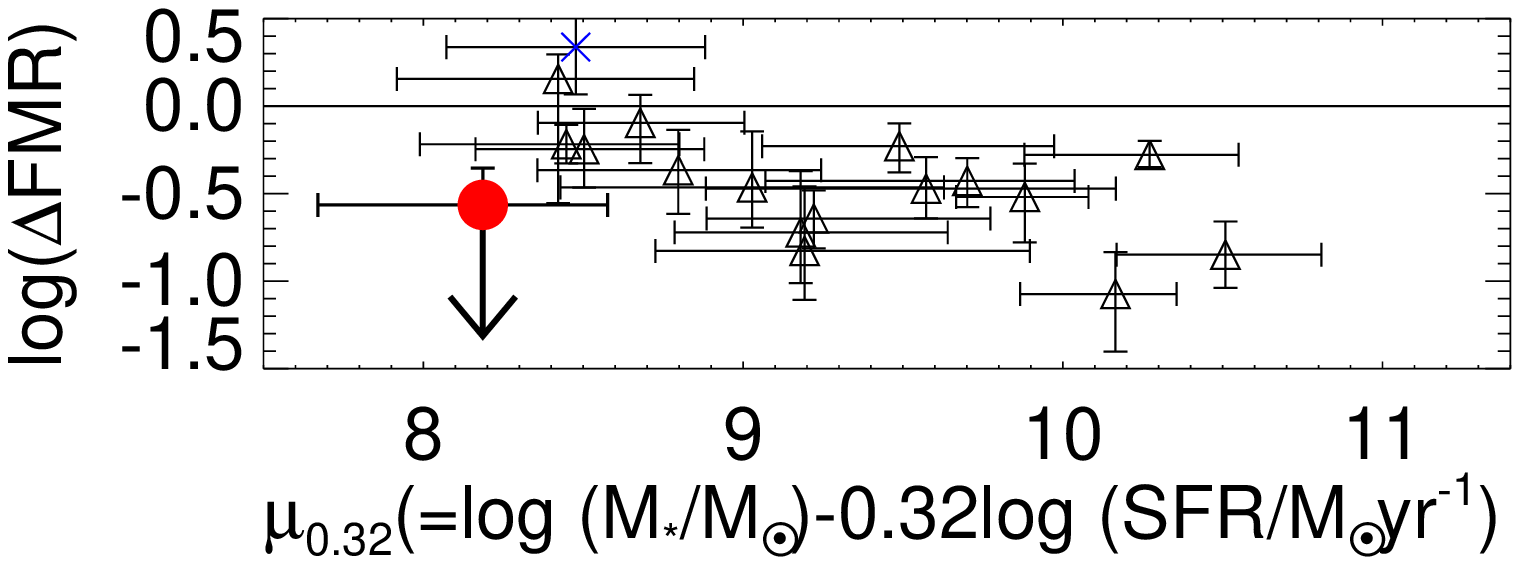}
 \caption{\textit{Upper panel:} Mass-metallicity relation at $z\geq$\,3.   \textit{Bottom panel:} deviation {in metallicity ($\Delta$\,FMR$=$\,$Z_{\rm obs}-Z_{\rm FMR}$) from} the FMR  {\citep{Mannucci2011}} as a function of $\mu_{32}$. {Dotted, dashed, dot-dashed and solid lines in the upper panel show the MZR at $z$\,$\sim$\,0.07, 0.7, 2.2 and 3.4 presented by \citet{Troncoso2013}. Solid line in the lower panel indicates a perfect match between measured metallicities and the ones derived through the FMR. } 
 LBGs from \citet[][triangles]{Troncoso2013} and the low-mass lensed galaxy "The Sextet" at $z=3.04$ from \citet[][blue cross]{Belli2013} are shown for comparison.} 
   \label{fig:MZR}
   \end{figure}

{In Figure~\ref{fig:MZR} we use our metallicity limit and the 
stellar mass\footnote{{Stellar masses have been derived following  \citet{Finkelstein2012} by fitting models accounting for nebular 
(line plus continuum) emission to the lens-subtracted SED of the source in 
four HST bands, which correspond to rest-frame UV. Photometry at longer wavelengths (optical rest-frame) were not included as source plus lens 
emission could not be deblended. 
However, we note that the sum of the best-model luminosities in the optical (rest-frame) for lens and source appears consistent, being only slightly 
lower than the total observed luminosities after de-magnification 
(see Fig.3 in vdWel13).}}
derived by vdWel13 to show the good agreement found 
between the position of $J1000+0221S$ and the extrapolation to 
low stellar masses of the MZR traced by more massive LBGs at 
$z \ga$\,3.}
Compared to the other few galaxies of similar or slightly higher 
masses and redshift, the upper limit in metallicity of $J1000+0221S$ is 
$\sim$\,0.5 dex lower. 

The scatter and normalization of the MZR at low-$z$ have  
been associated to the star formation activity and to the 
presence of intense gas flows in a tight relation between mass, 
metallicity and SFR, the so-called ``Fundamental Metallicity Relation'' 
\citep[FMR, ][]{Mannucci2010}. 
According to the FMR, at a given stellar mass, galaxies with higher 
SFR do have lower metallicities 
\citep[see also][]{Perez-Montero2013}. 
In contrast to the MZR, the FMR has been found to persist  
in galaxies out to redshift $z \sim$\,2.5
\citep[][]{Mannucci2011,Belli2013}. 
However, at $z\ga$\,3 most LBGs studied so far, e.g. those in the
AMAZE and LSD samples, are found to be more metal-poor 
than predicted by the FMR. This may suggest a change in the 
mechanisms giving origin to the FMR or a strong selection 
effect at these redshifts \citep{Troncoso2013}. 
Alternatively, it may suggest 
that metallicity calibrations based on local galaxies  
could not apply to high-$z$ galaxies due to their comparatively  
higher excitation/ionization conditions \citep{Kewley2013}. 
Also, recent studies on local galaxies using integral field spectroscopy 
have questioned the validity of the FMR as due to aperture biases  \citep{Sanchez2013}. 

In Figure.~\ref{fig:MZR} we reproduce the  
results by \citet{Troncoso2013} {for the FMR} and 
include the position of $J1000+0221S$ 
{using the extrapolation of the FMR 
to low stellar masses by \citet{Mannucci2011}.} 
Clearly, the metallicity of $J1000+0221S$ is at least 0.55 dex 
lower than predicted by the FMR. 

The very low metallicity, high specific SFR ($\sim$\,10$^{-8}$\,yr$^{-1}$) 
and extremely high EWs of $J1000+0221S$ are indicative 
of a rapidly growing galaxy in an early stage of its evolution. 
The offset position found in the MZR and FMR by at least 1 dex 
and 0.5 dex, respectively, compared to the local relations 
suggests the action of massive gas flows 
\citep[e.g.][]{Dayal2013}. 
One interesting possibility is that the recent star 
formation in $J1000+0221S$ is being fed by 
massive accretion of pristine gas in the cold-gas flows 
mode, as predicted by cosmological simulations as the 
main mode of build-up galaxies at these redshifts  
\citep[e.g.][]{Dekel2009} and supported by 
observational evidences in some low-metallicity 
starbursts \citep{Cresci2010,Sanchez-Almeida2014}. 

\subsection{The first estimation of Lyman Continuum escape fraction in
the very low luminosity regime $L=0.04 L^{\ast}$}

The discovery of $J1000+0221S$ also offer the great
opportunity to derive one of the first limits to the
LyC escape fraction at $z>3$ in a regime of very low
intrinsic luminosity ($L \sim$\, 0.05$L^{*}$), as 
suggested by \citet{vanzella12}.
The photometry of this object has been derived
by \citet{boutsia13} from deep LBC data in the $UGR$ filters used
in that work to search for LBGs at $2.7\le z\le 3.4$.
This lensed galaxy has AB magnitudes $R=24.15\pm 0.02$, 
$G=25.22\pm 0.04$ and an upper limit $U\ge 28.93$ at 1$\sigma$.
Note that this galaxy was not selected as an LBG candidate 
in \citet{boutsia13} due to its color $G-R=1.07$, which is 
slightly redder than the typical color cut $G-R\le 1$ adopted. 
This is due to the fact that, as shown in vdW13, the SED of this
galaxy is contaminated by the lens galaxy, an elliptical at $z=1.53$.

Using the LBC photometry alone, we can derive a limit to the LyC 
escape fraction for the lensed galaxy, adopting the same
technique already used in \citet{boutsia11}. To get rid of the light
contamination by the elliptical $z=1.53$, we have checked the photometry
in the ACS band $V_{606}$, which is the closest HST band to our $R$
filter. The total magnitude of the lens+source system is $V_{606}=24.15$, equal
to our $R$ band magnitude. Thus we can safely assume that the contribution
by the lens is corresponding to $R=26.4$ and the corrected magnitude
of $J1000+0221S$ is $R=24.3$.
Adopting the upper limit in the $U$ band as an estimate for the maximum
flux emitted by the lensed source and using the corrected $R$ band flux, we
 derive the relative escape fraction simply by
\begin{equation}
f^{rel}_{esc}=\frac{L_{1500}/L_{900}}{flux_R/flux_U}exp(\tau_{IGM}) 
\end{equation}

As in \cite{boutsia11}, we adopt a value of 3 as an estimate for the
intrinsic ratio $L_{1500}/L_{900}$. {Following \citet{Prochaska2009}
we derive a correction for the IGM transmission of
$\exp(-\tau_{IGM})=0.1811$ at the redshift of the source, $z=3.417$.}
We thus obtain an upper limit to the escape fraction of 23.2\% at 1$\sigma$
confidence level. While this limit is not so stringent with respect to
other estimates in the literature at z=3 \citep[e.g.][]{boutsia11,vanzella10,
mostardi13}, it is nonetheless important since we are probing an intrinsic
luminosity regime still unexplored before this work.
The source magnitude corrected for lensing is $R=28.3$, corresponding to
an absolute magnitude of $-17.4$ or equivalently to $L_{1500}=0.036
L^{\ast}(z=3)$, assuming a typical value $L^{\ast}(z=3)=-21.0$.
This is equivalent to a depth of 32.9 magnitudes in the $U$ band, after
correcting our upper limit for the magnification factor.

A number of theoretical models \citep[e.g.][]{Nakajima2013,paard13,ferrara13,Dijkstra2014} 
are investigating the processes involved during the end of the
so-called ``Dark Ages''. 
The main-stream under all these models is that the sources 
responsible for re-ionizing the Universe are dwarf galaxies
(sub-$L^{\ast}$) at $z\sim z_{reion}\sim 7$. 
Some of these works are also assuming a LyC escape fractions 
greater than 30-50\% even at lower redshifts 
\citep[e.g.][]{Nakajima2013,Dijkstra2014}. 

Due to IGM absorption increasing with redshift, the LyC escape
fraction can only be directly measured up to $z\sim 3$.  
For this reason our limit on the $f_{\rm esc}$ from an ultra-faint 
$z=3.4$ galaxy provides an interesting input to reionization 
model predictions, under the assumption that faint galaxies
such as $J1000+0221S$ are representative of the whole 
faint galaxy population at $z>3$. 
In fact, our stringent 23\% limit is at the lowest boundary of the 
range $f_{\rm esc}\sim20-30$\% which, according to recent
observational evidence \citep{Finkelstein2012,Grazian2012}, allows 
star-forming galaxies to keep the IGM ionized at $z>6$. 

Our current results for $J1000+0221S$ highlight the 
strength of strong lensing techniques to study the properties of low-mass 
star-forming galaxies at $z\ga$\,3. 
If, as expected, analogs of $J1000+0221S$ are ubiquitous at  
higher redshifts, forthcoming data from the CANDELS 
survey and the HST Frontier Fields Initiative and its spectroscopic 
follow-ups, will likely provide a statistically significant number of 
such systems, needed to derive more stringent limits to the 
escape fraction of LyC photons by ultra-faint galaxies and 
to study in larger detail the mechanisms driving the early phases 
of galaxy formation.

\acknowledgments
We wish to thank Roberto Maiolino for helpful assistance with metallicity determinations.  
We thank the referee for helpful suggestions that contribute to improve the original manuscript. This work is partially founded by the FP7 SPACE project “ASTRODEEP” (Ref.No: 312725), supported by the European Commission. 



{\it Facilities:}\facility{LBT (LUCI)}, \facility{LBT (LBC)}, \facility{HST (ACS, WFC3)}, \facility{KPNO (NEWFIRM)}.

\end{document}